\documentclass[aps
 ,prb
 ,amsmath,amssymb
 ,twocolumn
 ,floatfix
 ,reprint
 ,superscriptaddress
]{revtex4-2}
\usepackage[english]{babel}
\usepackage{graphicx}
\usepackage{placeins} 
\usepackage{enumitem} 
\usepackage{dcolumn}
\usepackage{bm}
\usepackage{gensymb} 
\usepackage{xfrac}
\UseCollection{xfrac}{plainmath}

\usepackage[normalem]{ulem}
\usepackage{xcolor}
\definecolor{eloicyan}{rgb}{0.,0.64,0.84}
\definecolor{eloired}{rgb}{0.8.,0.64,0.1}
\definecolor{darkgreen}{rgb}{0.,0.46,0.46}
\definecolor{lighgreen}{rgb}{0.2,0.75,0.56}
\definecolor{darkblue}{rgb}{0.1,0.2,0.46}
\definecolor{lightred}{rgb}{1.0,0.3,0.3}
\definecolor{darkred}{rgb}{.8,0.1,0.3}
\definecolor{lilac}{rgb}{0.6,0.2,0.7}

\renewcommand{\vec}[1]{\boldsymbol{\mathbf{#1}}}

\DeclareMathAlphabet\mathbfcal{OMS}{cmsy}{b}{n} 
\DeclareUnicodeCharacter{2212}{-} 

\begin{document}


\title{Statics and Dynamics of Skyrmions in Balanced and Unbalanced Synthetic Antiferromagnets}

\author{Eloi~Haltz}
\email{eloi.haltz@univ-paris13.fr}
\affiliation{School of Physics and Astronomy, University of Leeds, Leeds LS2 9JT, United Kingdom}
\affiliation{LSPM-CNRS, UPR3407, Sorbonne Paris Nord University, Villetaneuse, France}

\author{Christopher~E.~A.~Barker}
\altaffiliation[Present address: ]{National Physical Laboratory, Hampton Road, Teddington TW11 0LW, United Kingdom}
\affiliation{School of Physics and Astronomy, University of Leeds, Leeds LS2 9JT, United Kingdom}

\author{Christopher~H.~Marrows}
\email{c.h.marrows@leeds.ac.uk}
\affiliation{School of Physics and Astronomy, University of Leeds, Leeds LS2 9JT, United Kingdom}

\date{\today}

\begin{abstract}
Synthetic antiferromagnets have great potential as skyrmion carriers in which new properties are expected for these spin textures, owing to changed magnetostatics and the absence of net topological charge. Here we numerically simulate the static and dynamic behaviour of skyrmions in these systems and clearly highlight the benefits compared to ferromagnetic single layers. In particular, our results show a reduction of the skyrmion radius, an increase of their velocity under current, and a vanishing of their topological deflection. We also provide a robust and straightforward analytical model that captures the physics of such skyrmions. Finally, by extending the model to the case of an unbalanced SAF, we show some conditions for the system that optimise the properties of the skyrmion for potential spintronic devices.
\end{abstract}

\keywords{Suggested keywords}
\maketitle

\section{Introduction}

Magnetic skyrmions are topological magnetic textures with a core magnetisation pointing in the opposite direction to the surrounding magnetisation~\cite{Fert2013,Nagaosa2013,Back2020}. Their predicted small size, topological stability, and ease of manipulation with spin-torques makes them promising candidates as information carriers for future technology~\cite{Fert2017,Bourianoff2018,Zhang2020,Li2021,Marrows21,Vakili2021}. In the last decade, theoretical models, numerical simulations, and experiments furthered  understanding of their properties~\cite{Everschor-Sitte2018,Back2020,Buttner2018}. However, some key experimental barriers such as reducing their size, enhancing their stability at room temperature, or their low-power manipulation under electrical current have yet to be overcome, meaning that they are rarely integrated in actual devices~\cite{Back2020,Vakili2021}.

To surmount these limitations, it has been proposed to consider these magnetic textures in antiferromagnetic (AF) systems instead of the ferromagnetic single layers (SL)~\cite{Barker2016,Zhang2016,Buttner2018,Bhukta2022}. By switching to these multi-sublattice systems, two crucial points are addressed. First, there is a drastic reduction of stray fields due to the lack of a net magnetisation, which should reduce the size of the skyrmions. Second, we can expect a cancellation of the topological deflection of the skyrmions present in each of the anti-aligned magnetic lattices that allows them to be driven along the direction of the applied current~\cite{Barker2016,Zhang2016c,Buttner2018}. One of the promising systems for the stabilisation of such antiferromagnetic skyrmions are synthetic antiferromagnets (SAFs)~\cite{Parkin1991}. In these systems, multiple ferromagnetic layers are antiferromagnetically coupled through non-magnetic spacer layers by means of the Ruderman-Kittel-Kasuya-Yosida (RKKY)-like indirect exchange interaction~\cite{Zhang2016c,Duine2018}. Even if skyrmions have recently been experimentally observed in a SAF~\cite{Dohi2019,Legrand2020,Finco2021,Juge2022}, a clear and simple micromagnetic description of both the statics and dynamics of such skyrmions is so far lacking.

In order to improve the description of these properties and highlight the benefits of using a SAF over conventional ferromagnetic SLs, we numerically simulate the behaviour of skyrmions in a bilayer SAF and their dependence on a large range of parameters. We also adapt the ferromagnetic analytical formalism of skyrmion stability~\cite{BernandMantel2018} and dynamics~\cite{Thiele1973} to SAF systems. This leads to results that are in good agreement with the numerical simulations whilst also clarifying the underlying mechanisms. First, we investigate the stability and the size of SAF skyrmions and how their radii evolve with micromagnetic parameters. Second, we study the dynamics of these skyrmions under spin currents. Finally, we unbalance the SAF skyrmion by either making the magnetic layers constituting the SAF asymmetric, or by applying an external field which have different effect on each layers in order to verify the conditions for their enhanced velocity and vanishing skyrmion Hall angle. The results obtained qualitatively show the benefits of SAF skyrmions and propose some methods for more quantitative optimisation.

\section{Numerical simulations}

The stability and the behaviours of magnetic skyrmions have been numerically simulated by using the micromagnetic \textsc{mumax$^3$} software~\cite{Vansteenkiste2014}. We have studied both pairs of antiferromagnetically coupled layers (SAFs), as well as ferromagnetic single layers (SLs) for comparison. For each situation that we simulate, the magnetic state is initialised with a random skyrmion-like texture and relaxed before any eventual current injection. For the ferromagnetic cases, a single level mesh ($512 \times 512 \times 1$ of cubic cells of side 1~nm) is considered. For the SAF cases, a stack of three layers with the same dimensions is considered: two magnetic (indexed 1 and 2) separated by a non-magnetic spacer layer. The RKKY-like indirect exchange coupling between the two magnetic layers is accounted for as a space-dependent field $H_\mathrm{RKKY}(x,y,1)=\frac{-J_{12}/t_1}{\mu_0 M_1} \vec{m}(x,y,2)$ acting on the normalised magnetic moment $\vec{m}(x,y,2)$ in layer 2 (the bottom layer) and $H_\mathrm{RKKY}(x,y,2)$ (with 1$\leftrightarrow$2) acting on the normalised magnetic moment $\vec{m}(x,y,1)$ in layer 1 (the top layer). $|J_{12}|=1\times10^{-3}$~J/m$^2$ is the RKKY coupling parameter (constant for all the presented results), $t_1$ and $t_2$ and $M_1$ and $M_2$ are the thickness and the magnetisation of each of the magnetic layers. For all the following plots, the red and blue points correspond to the values for the top and the bottom magnetic layer (indexed 1 and 2, respectively) of the SAF and the grey points correspond to the isolated ferromagnetic SL.

\section{Results and discussion}

\subsection{Phase diagram}

\begin{figure}[t]
    \includegraphics[width=8cm]{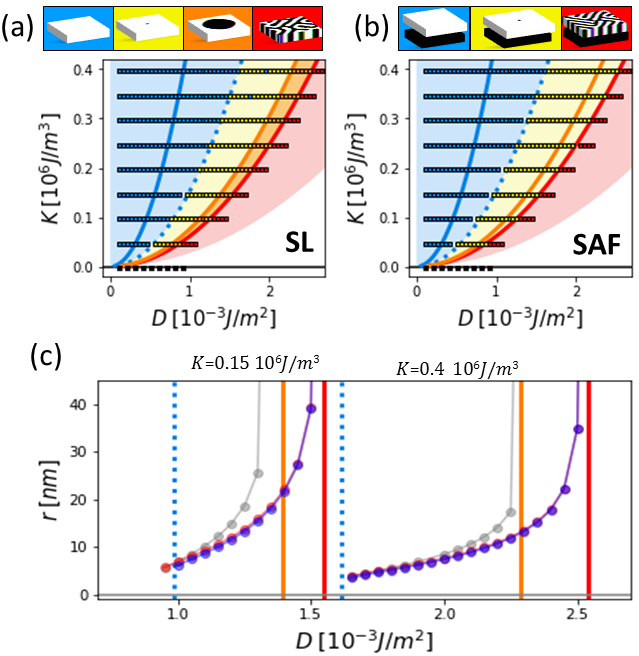}
    \caption{Static properties of a skyrmion in a SAF by comparison to an isolated SL: (a) and (b) Phase diagrams of magnetic textures resulting for the relaxation of a skyrmion in a SL (a) and in a SAF (b) for different anisotropy and DMI parameters $D$ and $K=K_0-\frac{\mu_0}{2}M_\mathrm{s}^2$. The top panel shows a sketch of the obtained stable magnetic textures for all the simulated cases (indicated as squares in the main plot). The blue, orange, and red lines correspond to the expected phases boundaries for the saturated/skyrmion, skyrmion/bubble and bubble/maze state in a magnetic SL according to the reference~\cite{BernandMantel2018}. The blue dotted line is a guide to the eye $\propto\sqrt{AK}$ that shows the artificial saturated/skyrmion limit due to the discretization of the micromagnetic simulation. (c) Evolution of the skyrmion radius \textit{versus} $D$ for a skyrmion in a SL (in grey) and in a SAF (in red and blue for each of the skyrmions in the two SAF layers) for two different values of effective anisotropy $K$ (indicated in the main plot). The vertical lines correspond to the phase boundaries (as for (a) and (b)) for theses two values of $K$. These results have been obtained for $M_\mathrm{s} = M_1 = M_2 =0.8 \times 10^6$~A/m and $A = A_1 = A_2 = 10 \times 10^{-12}$~J/m.
    \label{fig:Fig1}}
\end{figure}

First, to find the parameters at which the skyrmions are stable, we calculated the magnetic textures resulting from the relaxation of a skyrmion-like texture in a SL and in a SAF. Fig.~\ref{fig:Fig1}(a) shows the phase diagram obtained in a ferromagnetic SL for different Dzyaloshinkii-Moriya interaction (DMI) strength $D$ and perpendicular magnetic anisotropy (PMA) $K$, with a square marking the position of each calculation. Here, $K$ refers to the effective anisotropy : $K=K_0-\frac{\mu_0}{2}M_\mathrm{s}^2$ where $K_0$ is the PMA induced by the interfaces, which competes against the demagnetisation effect induced by the magnetisation $M_\mathrm{s}$. A magnetisation of $M_\mathrm{s}=0.8 \times 10^6$~A/m and an exchange stiffness of $A = 10 \times 10^{-12}$~A/m$^2$ are considered.

Four distinct magnetic textures are obtained as sketched in the top panel of Fig.~\ref{fig:Fig1}(a). For an increasing DMI, there are: the saturated uniform magnetisation (in blue), skyrmions with small radius (in yellow), magnetic bubbles with much larger radius (in orange), and a maze state (in red). The origin of these four phases is well described by the usual approaches~\cite{Wang2018b,BernandMantel2018}. The full lines show the expected boundaries between these phases. The horizontal black line corresponds to an out-of-plane easy axis for $K>0$. The saturated/skyrmion phase boundary corresponds to the critical DMI parameter $D=\left( \frac{4}{\pi}-\frac{8}{\pi^2}\right) \sqrt{AK}$ shown as a blue line. The discrepancy of that boundary is due to the finite size of the mesh which cannot handle the stabilization of skyrmions with sizes too close to the cell dimensions~\cite{Wang2018b}. The blue dotted line is a guide for the eye ($\propto\sqrt{AK}$) that follow that limit. The orange line corresponds to the skyrmion/bubble transition~\cite{BernandMantel2018}. The maze state proliferation corresponds to the critical DMI~\cite{BernandMantel2018,You2015,Rohart2013} of $D=\frac{4}{\pi} \sqrt{AK}$.

Fig.~\ref{fig:Fig1}(b) shows the phase diagram obtained in a similar way but for skyrmions in a balanced SAF stack, i.e. one in which the values of $K$ and $D$ are varied in the same way in both layers. Even if the global shape is similar, only three types of magnetic textures are visible: for an increasing DMI, the saturated state (in blue), the skyrmion state (in yellow) and, the maze state (in red). The bubble state is no longer present. The plotted lines are the same as Fig.~\ref{fig:Fig1} (a) and seem to match the phases boundaries for the SAF except for the skyrmion/bubble one that no longer exists.

Fig.~\ref{fig:Fig1} (c) shows the variation of the skyrmion radius as a function of the DMI for two different values of effective anisotropy in a SL (in grey) and in a SAF (in red and blue for each of the skyrmions in the two SAF layers). For both cases, the radius increases with the DMI and diverges at the bubble or the maze transition for, respectively, the SL and the SAF. In general, the skyrmions are slightly smaller in the SAF by comparison with the SL. However, for small skyrmions, the radii are very similar in both systems.

To determine the stability of the different magnetic phases, it is possible to calculate the magnetic energy density of the SAF by integrating the different energetic contributions through the full depth of the SAF stack. 
As proposed~\cite{Haltz2021}, for large antiferromagnetic coupling, the RKKY energetic contribution is constant and it is possible to associate an effective ferromagnetic single layer to the stack with effective magnetisation, exchange stiffness, DMI, and anisotropy parameters : $M_\mathrm{s}=\frac{\sum (-1)^i M_i t_i}{\sum t_i}$, $A=\frac{\sum A_i t_i}{\sum t_i}$, $D=\frac{\sum D_i t_i}{\sum t_i}$, $K=\frac{\sum K_{0i} t_i}{\sum t_i}-\frac{\mu_0}{2} \frac{\sum M_{i}^2 t_i}{\sum t_i} $. Here, $M_i$, $A_i$, $D_i$ and $K_{0i}$ are the parameters of each magnetic layer $i$ composing the SAF stack. Thus, if the layers are equal, the effective parameters are the ones of just one of the layers that constitute the SAF, with the exception of the net magnetisation $M_\mathrm{s}$, which vanishes.

In that case, the SAF energy density only differs from the isolated magnetic layer by the long range demagnetisation effect also called flux closure. For skyrmions, that contribution increases with the magnetisation and with the skyrmion radius~\cite{Buttner2018,BernandMantel2018,You2015}. In a magnetic SL, it is responsible for the skyrmion/bubble transition~\cite{BernandMantel2018} as observed Fig.~\ref{fig:Fig1}(a) and (c). On the other hand, for the SAF, since the net magnetisation vanishes, that contribution disappears and the magnetic bubble phase is not stable anymore, as observed in Fig.~\ref{fig:Fig1}(b) and (c). Otherwise, since the effective parameters of the SAF are the same as those of a SL, the phase diagrams are similar for the two systems. The long-range effect tends to increase the skyrmion radius in the SL. However, for small skyrmions, that effect diminishes and the skyrmion radii are similar in both systems as shown in Fig.~\ref{fig:Fig1}(c). For bigger skyrmions, that contribution increases only for the SL, which makes the skyrmions larger compare to the SAF where that contribution remains zero. The vanishing of that long-range demagnetisation effect increases significantly the skyrmion stability region in the SAF in comparison to an isolated SL by removing the bubble phase as shown Fig.~\ref{fig:Fig1}(b) and (c).

In AF systems where the two anti-aligned magnetic sublattices are merged, such as pure AF or compensated ferrimagnets, the situation differs from the the SAF where the two anti-aligned magnetic sublattices are spatially separated. In both cases, the long-range effect disappears due to the vanishing of the net magnetisation. However, in systems where the sublattices are merged, the short-range demagnetisation effect also decreases and the effective anisotropy constant becomes $K=\sum K_{0i} -\frac{\mu_0}{2} M_\mathrm{s}^2$. In those systems, the vanishing of that short-range demagnetisation when $ M_\mathrm{s}$ goes to zero drastically reduces the skyrmion sizes compared to a SL. That is not the case in SAFs where even the skyrmions are only slightly smaller. That explains why in ferrimagnets the observed skyrmions are smaller that in a SAF where sizes are comparable to conventional ferromagnetic layers~\cite{Caretta2018,Legrand2020}.

\subsection{Dynamics under current}

\begin{figure}[t]
    \includegraphics[width=8cm]{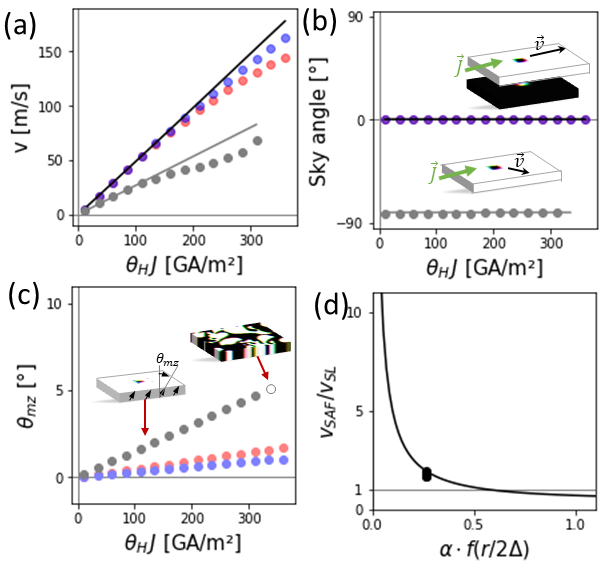}
    \caption{
    Dynamics and stability of a skyrmion under SOT in a SAF by comparison to an isolated SL. Skyrmion (a) velocity and (b) deflection \textit{versus} the spin-current density $\theta_\mathrm{SH} J$ for a skyrmion in a SL (in grey) and in the SAF (in blue and red). The points corresponds to the numerical simulations and the lines corresponds to the analytical model of $v$ in the two systems. When non-visible, the red and blue points or lines superimpose. The insets in (b) show a sketch of the skyrmion dynamics driven by SOT in both systems. (c) Angle of the magnetisation in the saturated region surrounding the skyrmion induced by the SOT \textit{versus} $\theta_\mathrm{SH} J$ as sketched in the inset. The second inset shows the destruction of the skyrmion in a single layer for large current. (d) Velocity ratio between a skyrmion in a SL and in a SAF $\sfrac{v_\mathrm{SAF}}{v_\mathrm{SL}}$ for different dissipation $\left(\alpha f\left(\frac{r}{2 \Delta}\right)\right)$. Points correspond to numerical simulations for $M_\mathrm{s} = 0.8 \times 10^6$~A/m, $A =10 \times 10^{-12}$~J/m, $K=0.7 \times 10^6$~J/m$^3$, $D=1.75 \times 10^{-3}$~J/m$^2$ and $\alpha=0.1$. The full line corresponds to the analytical model. 
    \label{fig:Fig2}}
\end{figure}

In this part, we investigate the benefits of moving from a magnetic SL to a SAF in term of skyrmion dynamics driven by spin orbit torque (SOT)~\cite{Back2020}. Fig.~\ref{fig:Fig2} shows the variations of the skyrmion velocity (a) and the skyrmion transverse deflection (b) \textit{versus} the spin-current density ($\theta_\mathrm{SH} J$) in a SAF (in red and blue) by comparing to an isolated SL (in grey). $J$ is the electrical charge current density and $\theta_\mathrm{SH}$ is the spin Hall angle. To compare the properties in both systems, the SOT is applied only in the top magnetic layer of the SAF stack, in a similar manner to previous modelling of domain wall dynamics~\cite{Alejos2018}. If the SOT were applied to both layers, the summation rules giving the effective parameters would lead to a current density twice as high as a SL case $\theta_\mathrm{SH} J = \frac{\sum \theta_\mathrm{H\,i} J_i t_i}{\sum t_i}$.

In both systems, the SOT induces a translational motion of the skyrmion with a velocity increasing with the current density. In both systems, the skyrmion velocity is linear with $J$ for small current density, but then deviates from linearity for increasing current. For the SL, the skyrmion is destroyed above a critical current density (here of 300~GA/m$^{-2}$), which fixes a maximum of velocity (of $\approx 55$~m/s for the considered parameters). For the SAF, the skyrmion is faster (almost twice for a given $J$). The skyrmions in each of the SAF layers start to separate for large current exceeding $\approx  200$~GA/m$^{-2}$ (for the considered value of RKKY-like coupling).

The destruction of the skyrmion is mainly due to the tilting of the magnetisation induced by the SOT. Fig.~\ref{fig:Fig2}(c) shows the average angle $\theta_{mz}$ of the magnetisation with the vertical axis in the saturated region surrounding the skyrmion. For both systems, $\theta_{mz}$ increases linearly with the current density. For the SL, that effect is in competition with the vertical easy-axis anisotropy leading to a linear dependence up to the reversal of the magnetisation (as shown in the insets Fig.~\ref{fig:Fig2}(c)). Before that critical value, the tilting also causes the $v$ curve to deviate from its expected linear variation with $J$. For the SAF, the SOT tilting is much smaller compared to the SL. That can be understood since this tilt results from the competition between the SOT, applied only in the top layer, and the anisotropy of each magnetic layer owing to the strong RKKY-like AF coupling. Thus, it is possible to apply much larger torques in the SAF before observing the non-linear regime and the destruction of the skyrmion. However, in addition to that tilt, the SOT will also tend to misalign the magnetization in each layer and separate the skyrmions in the two layers as shown Fig.~\ref{fig:Fig2}(a).

The deflection of skyrmions in both systems is sketched Fig.~\ref{fig:Fig2}(b). For the skyrmion in the isolated SL, the skyrmion is strongly deflected from the current direction by an angle of $\approx 85^\circ$ (the skyrmion Hall angle) that is constant with the current density. On the other hand, in the SAF the skyrmion moves along the current direction without any transverse deflection.

The velocity $\vec{v}$ of a spin texture driven by a force $\vec{\mathcal{F}}$ in a single ferromagnetic layer is well-described in the stationary regime by the Thiele equation~\cite{Thiele1973}:
\begin{equation}
    \vec{\mathcal{G}}\times\vec{v}-\alpha \left[\mathcal{D}\right]\cdot\vec{v}+\vec{\mathcal{F}}=0,
    \label{eq:Thiele}
\end{equation}
where $\vec{\mathcal{G}}$ and $\left[\mathcal{D}\right]$ are the gyrovector and the dissipation tensor coming from the precession and damping dynamics of the magnetic moments constituting the spin texture. For a skyrmion, $\vec{\mathcal{G}}=\pm \frac{\mu_0 M_s t}{\gamma_0} \vec{z}$ depending on the skyrmion core magnetisation direction $\vec{m}=\pm\vec{z}$ and $\left[\mathcal{D}\right]= \frac{\mu_0 M_s t}{\gamma_0} f\left( \frac{r}{2\Delta} \right) \mathbb{I}$ with $r$ the skyrmion radius and $\Delta\equiv\sqrt{A/K}$ the skyrmion wall width parameter~\cite{Rohart2013} ($\mathbb{I}$ is the identity matrix). $f(x)$ is a function linear in $x$ for large $x$ (i.e. $r\gg 2\Delta$) and which saturates to 1 for small $x$ ($r\approx 2\Delta$)~\cite{Rohart2013,Buttner2018}. In our case, we consider $f(x)\approx x+\frac{1}{1+x}$ to satisfy these two limits~\cite{Buttner2018,Panigrahy2022}.

When the SOT does not distort significantly the skyrmion structure, the resulting force on the skyrmion is proportional to the skyrmion radius and the spin current $\vec{\mathcal{F}}\propto r\, \theta_\mathrm{SH} \vec{J}$. That formalism gives some simple expressions for the skyrmion velocity and skyrmion deflection angle of:
\begin{equation}
    |\vec{v}|=\frac{\mathcal{F}}{\alpha \mathcal{D}} /\sqrt{1+\left(\frac{\mathcal{G}}{\alpha \mathcal{D}} \right)^2} \;\mathrm{ and }\; \frac{v_y}{v_x}=\frac{\mathcal{G}}{\alpha \mathcal{D}},
    \label{eq:vSk}
\end{equation}
where the film occupies the $x$-$y$ plane with the current flowing along the $x$-axis. These expressions reproduce with a good agreement what was obtained with the numerical simulations in a SL for low currents, as shown in Fig.~\ref{fig:Fig2}(a) and (b). The skyrmion radius is extracted from the micromagnetic simulations and used to calculate the velocity and the deflection with equation (\ref{eq:vSk}). When the current increases, the skyrmion structure is distorted by the SOT that lowers the resulting force, and then its velocity deviates from the linear behaviour. 

 In a SAF, the stationary dynamics of the skyrmion can be described by two Thiele equations~\cite{Zhang2016c}: one for each the skyrmions in the two layers with velocity $\vec{v}_1$ and $\vec{v}_2$ and parameters $\vec{\mathcal{G}}_1$ and $\vec{\mathcal{G}}_2$, $\alpha_1 \left[\mathcal{D}_1\right]$ and $\alpha_2 \left[\mathcal{D}_2\right]$, and $\vec{\mathcal{F}}_1$ and $\vec{\mathcal{F}}_2$. The RKKY coupling between the two layers is accounted as two additional forces exerted by the skyrmion in the layer 2 on the one in the layer 1 $\vec{\mathcal{F}}_{1\rightarrow2}$  and \textit{vice-versa} $\vec{\mathcal{F}}_{2\rightarrow1}$ with $\vec{\mathcal{F}}_{1\rightarrow2}=-\vec{\mathcal{F}}_{2\rightarrow1}$. If the two skyrmions are coupled, in the stationary regime $\vec{v}_1=\vec{v}_2=\vec{v}$. Thus, the two Thiele equations can be summed up to describe the dynamics of the system with a single Thiele equation with effective parameters $\mathcal{G}=\mathcal{G}_1+\mathcal{G}_2$, $\alpha\left[\mathcal{D}\right]=\alpha_1\left[\mathcal{D}_1\right]+\alpha_2\left[\mathcal{D}_2\right]$ and $\vec{\mathcal{F}}=\vec{\mathcal{F}}_1+\vec{\mathcal{F}}_2$ with $\vec{\mathcal{F}}_{1\rightarrow2}$ and $\vec{\mathcal{F}}_{2\rightarrow1}$ cancelling each other~\cite{Panigrahy2022}. To use these parameters in the equation~\ref{eq:vSk} reproduces the simulated results with a very good agreement for low currents as shown with full lines in Fig.~\ref{fig:Fig2}(a) and (b).

 If the two magnetic layers constituting the SAF are the same, since $\frac{M_1 t_1}{\gamma_{01}}=\frac{M_2 t_2}{\gamma_{02}}$, we have $\mathcal{G}_2=-\mathcal{G}_1$ and the net gyrovector cancels to zero: $\vec{\mathcal{G}}\rightarrow\vec{0}$. Also, if $\alpha_1\frac{M_1 t_1}{\gamma_{01}} f\left( \frac{r_1}{2\Delta_1} \right)=\alpha_2\frac{M_2 t_2}{\gamma_{02}} f\left( \frac{r_2}{2\Delta_2} \right)$, we have $\alpha_2\left[\mathcal{D}_2\right]=\alpha_1\left[\mathcal{D}_1\right]$ and the dissipation is double that compared to the SL: $\alpha\left[\mathcal{D}\right]\rightarrow2\alpha_1\left[\mathcal{D}_1\right]=2\alpha_2\left[\mathcal{D}_2\right]$. Here, the SOT is applied only in the top layer and the resulting force is unchanged compare to the SL: $\vec{\mathcal{F}}\rightarrow\vec{\mathcal{F}}_1$. In that case, the skyrmion deflection $\frac{v_y}{v_x}$ fully vanishes and then the skyrmion moves along the current direction. The skyrmion velocity in the SAF $v_\mathrm{SAF}$ is increased compared to the skyrmion in an SL $v_\mathrm{SL}$ by a factor $\frac{v_\mathrm{SAF}}{v_\mathrm{SL}}=\frac{1}{2}\sqrt{1+\left(\alpha f\left(\frac{r}{2 \Delta} \right) \right)^{-2}}$. That enhancement decreases with dissipation, i.e. for low $\alpha$ and small skyrmions. Fig.~\ref{fig:Fig2}(d) shows the evolution of that ratio with the dissipation $\left(\alpha f\left(\frac{r}{2 \Delta} \right) \right)$. The point corresponds to $v_\mathrm{SAF}/v_\mathrm{SL}$ for the velocities shown in Fig.~\ref{fig:Fig2}(a). This consideration gives an upper limit $\alpha f\left(\frac{r}{2 \Delta}\right) = \frac{1}{\sqrt{3}}$ above which the gain in velocity resulting from the cancellation of the gyrovector is counterbalanced by the rise of the dissipation  
 and so the skyrmion in the SAF is no longer any faster compared to the SL.

\subsection{Unbalanced SAF}

\begin{figure}[t]
    \includegraphics[width=8cm]{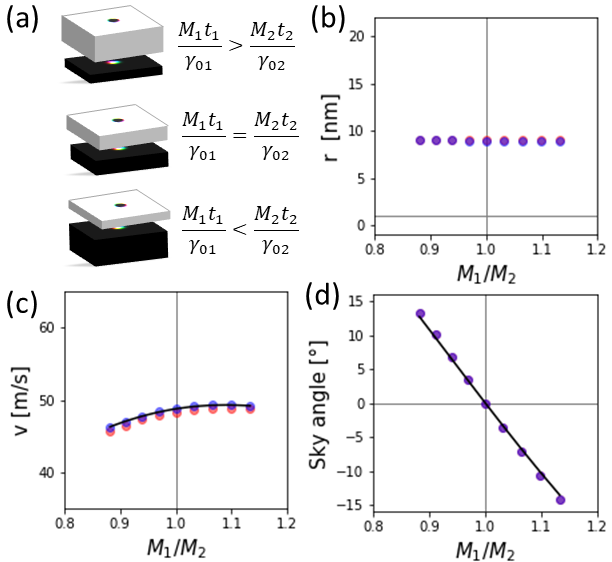}
    \caption{Behaviors of an angularly unbalanced SAF:
    (a) Sketch of the angularly unbalanced and balanced SAFs. (b) radius of the skyrmions in the top and bottom layers of the SAF \textit{versus} $R$ (in red and blue). (c) and (d) velocity and transverse deflection of the skyrmion \textit{versus} the unbalancing ratio $R=\frac{\mu_0 M_1 t_1}{\gamma_{01}}/\frac{\mu_0 M_2 t_2}{\gamma_{02}}=M_1/M_2$ here. The black line corresponds to the analytical model. The axis ranges for panels (b) and (c) are chosen to allow easy comparison with the equivalent panels in Fig.~\ref{fig:Fig4}.
    \label{fig:Fig3}}
\end{figure}

\begin{figure}[t]
    \includegraphics[width=8cm]{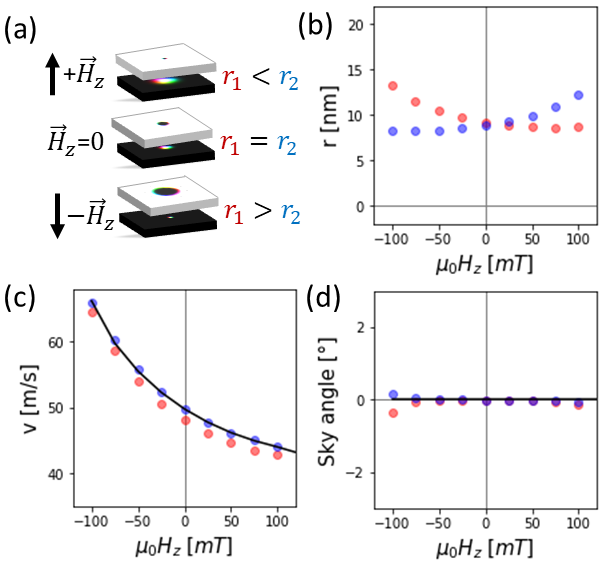}
    \caption{Behaviors of a geometrically unbalanced SAF:
    (a) Sketch of the geometrical unbalancing induced by an external field $H_z$. (b) Radius of the skyrmions in the top and bottom layers of the SAF (in red and blue) \textit{versus} $H_z$. (c) and (d) velocity and transverse deflection of the skyrmion \textit{versus} the out-of-plane field $H_z$. The black line corresponds to the analytical model.
    \label{fig:Fig4}}
\end{figure}

In this section, we investigate how the skyrmions behave in an unbalanced SAF, when the two magnetic layers are different. Two cases are possible. The first is when the stack is what we call angularly unbalanced so that $\frac{\mu_0 M_1 t_1}{\gamma_{01}}\neq\frac{\mu_0 M_2 t_2}{\gamma_{02}}$ (as shown Fig.~\ref{fig:Fig3}(a)) so that the layers differ in their angular momentum. The second is when the stack is said to be geometrically unbalanced $\frac{r_1}{2\Delta_1} \neq \frac{r_2}{2\Delta_2}$ (as shown Fig.~\ref{fig:Fig4} (a)), for instance as the result of the application of an external field to a balanced SAF. To simplify the discussion, we assume $\alpha_1=\alpha_2\rightarrow\alpha$ since different Gilbert damping parameters would induces similar behaviour to the geometrically unbalanced situation according to the definition of the net dissipation $\alpha\left[\mathcal{D}\right]=\alpha_1\left[\mathcal{D}_1\right]+\alpha_2\left[\mathcal{D}_2\right]$.

First, we study the effect of the angular unbalancing between the two magnetic layers, achieved in this case by modifying the magnetisation of each layer $M_1$ and $M_2$. In order to keep a quasi-constant radius for the skyrmions (as shown Fig.~\ref{fig:Fig3}(b)), we keep the quantity $M_1+M_2$ constant. The strong interlayer coupling $J_{12}$ means that we not see so much difference in skyrmion radius as in more weakly coupled systems~\cite{Lonsky2020}.  Fig.~\ref{fig:Fig3} shows the evolution of the skyrmion velocity (c) and the skyrmion deflection (d) \textit{versus} the unbalancing ratio ($R\equiv\frac{ M_1 t_1}{\gamma_{01}}/\frac{ M_2 t_2}{\gamma_{02}}$ with $R=M_1/M_2$ in our case) for a spin-orbit torque induced by a current density of 100~GA/m$^{-2}$ acting on the top layer. The effective Thiele parameters for the SAF reproduce with a very good agreement the simulated results as shown Fig.~\ref{fig:Fig3} (c) and (d).

Whenever $M_1\neq M_2$, the skyrmion deflections in the two layers do not cancel anymore and the skyrmion is deflected leading to a finite skyrmion Hall angle. This is directly associated with the fact that $\mathcal{G}\neq 0$ in equation (\ref{eq:Thiele}). For $M_1 > M_2$, $\mathcal{G}>0$ and the skyrmion is deflected in the same direction as for the SL (as shown Fig.~\ref{fig:Fig2}(b)). For $M_1 < M_2$ the deflection in the bottom layer is dominant and the skyrmion is deflected in the opposite direction. If the level of damping in each layer is close and the RKKY-like exchange is strong enough to keep the same shape for the two skyrmions in each layers, the skyrmion deflection evolves as $\frac{v_y}{v_x}=\frac{1}{\alpha f\left( r/2\Delta\right)} \frac{1-R}{1+R}$. Thus, the skyrmion with a small dissipation (\textit{i.e.} small $\alpha$ or small radius) will be much more sensitive to eventual angular unbalanced cases.

The skyrmion velocity continuously increases with $M_1/M_2$ and the balance point $M_1=M_2$ does not correspond to the fastest configuration as shown Fig.~\ref{fig:Fig3}(c). That is even true for the velocity along the current direction $v_x$ (not shown here). If we assume a constant radius for the skyrmion, the velocity ratio between the balanced and the unbalanced SAF is given by $\frac{v(R)}{v(R=1)}=\sfrac{\frac{2}{1+R}}{\sqrt{1+\left( \frac{1-R}{1+R}\right)^{2}\left( \alpha f\left( \frac{r}{2\Delta} \right)\right)^{-2}}} $. That velocity ratio exhibits a maximum for $R = \frac{1-\left( \alpha f\left( r/2\Delta \right)\right)^2}{1+\left( \alpha f\left( r/2\Delta \right)\right)^2}$ 
which deviates from the balanced case ($R=1$) as the dissipation increases. The lack of symmetry around the balanced configuration $R = 1$ is due to the fact that we are only exerting SOT on the top layer in our simulation.

These above results show that skyrmion deflection vanishes only if the angular momenta of each layer compensate each other \textit{i.e.} $\frac{ M_1 t_1}{\gamma_{01}}=\frac{ M_2 t_2}{\gamma_{02}}$. If this condition is not satisfied, the skyrmion is deflected and that deflection is even larger for low dissipation systems with small damping and small radius. It worth mentioning that for SAFs with TM-based FM layers, $\gamma_{01}\approx\gamma_{02}$, thus, whenever $M_1 t_1 \approx M_2t_2$, we have an angular balanced SAF and the skyrmion moves along the current direction. Furthermore, the quantity $|M_1 t_1 - M_2t_2|$ is exactly what is measured with usual magnetometry experiments (such as VSM or SQUID). On the other hand, the velocity is not maximum for the exactly balanced case. The value of $R$ at which the velocity maximum occurs deviates from the angular compensation as the dissipation increases.

Second, we study the effect of geometrical unbalancing between the two skyrmions in each magnetic layers where we can modify their radii $r_1$ and $r_2$ by means of an out-of-plane field $H_z$ (as shown Fig~\ref{fig:Fig4}(a)). In this approach we can make certain that none of the micromagnetic parameters differ in each magnetic layer.

Fig.~\ref{fig:Fig4}(b) shows how the radius of the skyrmion in each layer $r_1$ and $r_2$ evolve as the out-of-plane field $H_z$ is varied. For a positive $H_z$, the skyrmion in the upper layer (layer 1; with a down core magnetisation) shrinks while the skyrmion in the lower layer (layer 2; with an up core magnetisation) grows. Despite this difference between $r_1$ and $r_2$, the skyrmions of each layer remain coupled together in the considered range of the applied field.

Fig.~\ref{fig:Fig4} shows the evolution of the skyrmion velocity (c) and the skyrmion deflection (d) \textit{versus} the external field for a spin-orbit torque induced by a current density of 100~GA/m$^{2}$ acting on the top layer. The effective Thiele parameters for the SAF, plotted as a continuous line, reproduce with a very good agreement the numerically simulated results as shown Fig.~\ref{fig:Fig3} (c) and (d). Fig.~\ref{fig:Fig4}(c) shows that the skyrmion velocity increases with $r_1$ as expected since $v \propto \mathcal{F}\propto r_1$ as the current is applied only the the top layer. Also, the fact that $r_2$ slightly decreases makes the effective dissipation $\alpha \mathcal{D}$ smaller compare to the balanced case, which also increases the skyrmion velocity. Interestingly, even if the $r_1\neq r_2$, the skyrmion still moves along the current direction without any deflection as shown Fig.~\ref{fig:Fig4} (d). This is directly associated with the fact that $\mathcal{G}$ only depends on the chirality of the skyrmion in each layers and not on the skyrmion shape. Thus, whenever we have angular balance ($R=1$) the skyrmion moves without any deflection even if we have geometrical imbalance.

In experimental samples, the use of out-of plane fields~\cite{Herve2015} or bias field-like effect~\cite{Legrand2020} is usually applied to stabilise the skyrmion. Additionally, residual Oersted fields can also exist because of the large current injection of electrical currents~\cite{Zeissler2020}. These results demonstrate that these external fields will not affect directly the deflection motion of the skyrmions in SAF stacks as it does for SLs.

These results present the variation of the skyrmion sizes induced by external applied field but similar results can be obtained for different micromagnetic parameters such as interfacial terms ($K_{01}\neq K_{02}$ and $D_1\neq D_2$) or intrinsic terms ($A_1\neq A_2$) whenever they do not change the net angular momentum. Thus, it shows that the two magnetic layers constituting the SAF do not have to be constituted of the same material whenever $\frac{ M_1 t_1}{\gamma_{01}}=\frac{ M_2 t_2}{\gamma_{02}}$. Also, when the current is injected only in one layer, it is better to have the skyrmion in this one as big as possible to increase the net force $\vec{\mathcal{F}}=\vec{\mathcal{F}_1}\propto r_1$ and the one in the other layer as small as possible to decrease the net dissipation $\alpha \mathcal{D}$ which depends roughly on $\sim r_1+r_2$.

\section{Conclusion}

In this study we have investigated the static and the dynamic properties of skyrmions in SAF stacks by the mean of numerical micromagnetic simulations. First we have compared the properties of these systems with those of the usual ferromagnetic SLs to highlight their benefits in terms of skyrmion properties. We have shown a larger parameter range of stability and a quite smaller sizes for skyrmion in SAFs compared to SLs. We have also studied their dynamics under SOT and we have shown a vanishing of the skyrmion deflection and increases of their velocity in SAFs. By considering and effective analytical model based on Thiele equation, we have been able to reproduce and describe these obtained results. This model also highlight the relevant quantities that govern these skyrmion dynamics: the net angular momentum witch mainly governs the skyrmion deflection and the dissipation (depending on the magnetic damping and on the skyrmion radius) which mainly governs the skyrmion velocity. In particular we have shown that the cancellation of the net angular momentum is directly responsible for the deflection vanishing. We have also shown that the skyrmions in a SAF are faster compared to ferromagnetic SL only for a dissipation below a certain limit.

The model we have developed also allows the description of the skyrmion properties in an unbalanced SAF, when the layers constituting the stack are different. This have shown that the topological deflection vanishes only if the SAF stack is angularly balanced: if the angular momenta of each layer compensate each other: $\sum \frac{ (-1)^i M_i t_i}{\gamma_{0i}}=0$. However, all the other micromagnetic parameters, even the skyrmion radii, do not affect these properties. These results show the possibility to differentiate the different layers constituting the SAF stack or to use an out-of-plane bias field in order to optimise the skyrmion velocities. The results and the simple model developed here can be a good base for further optimisations of real SAF stacks.

\begin{acknowledgements}
We acknowledge fruitful discussions with J.Barker, Kayla Fallon, Stephen McVitie, and Yves Roussign\'{e}. This work was supported by the EPSRC, grant number EP/T006803/1. C.E.A.B. acknowledges support from the National Physical Laboratory. A part of these numerical simulations was performed on MAGI, the computing platform of the University Sorbonne Paris Nord, managed by Nicolas Greneche.
\end{acknowledgements}

\bibliography{main.bib}

\end{document}